\documentclass[pre,superscriptaddress,twocolumn]{revtex4-1}
\usepackage{amsmath}
\usepackage{color}
\usepackage{amssymb} 
\usepackage{amsfonts}   
\usepackage{graphicx}
\graphicspath{{./Figures/}}
\usepackage{xcolor}
\usepackage{nicefrac}
\usepackage{url}
\usepackage{hyperref}

\newcommand{\beq}{\begin{equation}}
\newcommand{\eeq}{\end{equation}}

\newcommand{\thomas}[1]{{#1}}

\newcommand{\<}{\langle}
\renewcommand{\>}{\rangle}
\newcommand{\mI}{\mathcal{I}}
\newcommand{\mS}{\mathcal{S}}

\newcommand{\OEB}{Department of Organismic and Evolutionary Biology, Harvard University, Cambridge, MA, USA}
\newcommand{\LPENS}{Laboratoire de physique de l'\'Ecole Normale Sup\'erieure, CNRS, Sorbonne Universit\'e, Universit\'e de Paris, and \'Ecole normale sup\'erieure (PSL), 24 rue Lhomond, 75005 Paris, France}
\newcommand{\MoscowSOIBC}{Shemyakin-Ovchinnikov Institute of Bioorganic Chemistry, Moscow, Russia}
\newcommand{\MoscowPRNRMU}{Pirogov Russian National Research Medical University, Moscow, Russia}
\newcommand{\lastequal}{Corresponding authors. These authors contributed
  equally.}

\begin{document}

\title{Immune Fingerprinting through Repertoire Similarity}

\author{Thomas Dupic}
\affiliation{\OEB}
\author{Meriem Bensouda Koraichi}
\affiliation{\LPENS}
\author{Anastasia Minervina}
\affiliation{\MoscowSOIBC}
\author{Mikhail Pogorelyy}
\affiliation{\MoscowSOIBC}
\affiliation{\MoscowPRNRMU}
\author{Thierry Mora}
\thanks{\lastequal}
\affiliation{\LPENS}
\author{Aleksandra M. Walczak}
\thanks{\lastequal}
\affiliation{\LPENS}

\begin{abstract}
 % abstract
Immune repertoires provide a unique fingerprint reflecting the immune history of individuals, with potential applications in precision medicine. However, the question of how personal that information is and how it can be used to identify individuals has not been explored. Here, we show that individuals can be uniquely identified from repertoires of just a few thousands lymphocytes. We present ``Immprint,'' a classifier using an information-theoretic measure of repertoire similarity to distinguish pairs of repertoire samples coming from the same versus different individuals. \thomas{Using published T-cell receptor repertoires and statistical modeling, we tested its ability to identify individuals with great accuracy, including identical twins, by computing false positive and false negative rates $<10^{-6}$ from samples composed of 10,000 T-cells. We verified through longitudinal datasets and simulations that the method is robust to acute infections and the passage of time.} These results emphasize the private and personal nature of repertoire data.

\end{abstract}

\maketitle

 % maintext
\section*{Introduction}

Personalized medicine is a frequent promise of next-generation sequencing. These high-throughput and low-cost sequencing technologies hold the potential of tailored treatment for each individual.
However, progress comes with privacy concerns. Genome sequences cannot be anonymized: a genetic fingerprint is in itself enough to fully identify an individual, with the rare exception of monozygotic twins. The privacy risks brought by these pseudonymized genomes have been highlighted by multiple studies \cite{Homer2008, NAVEED2015,Sweeney2013},
and the approach is now routinely used by law enforcement. Sequencing experiments that focus on a limited number of expressed genes should be less prone to these concerns. However, as we will show, B- and T-cell receptor (BCR and TCR) genes are an exception to this rule.

BCR and TCR are randomly generated through somatic recombination \cite{Hozumi:1976p5097}, and the fate of each B- or T-cell clone depends on the environment and immune history. The immune T-cell repertoire, defined as the set of TCR expressed in an individual, has been hailed a faithful, personalized medical record, and repertoire sequencing (RepSeq) as a potential tool of choice in personalized medicine \cite{Robins2013,attaf_2015,Woodsworth2013,Bradley2019,Davis2019}. \thomas{In this report, we describe how, from small quantities of blood (blood spot or heel prick), one can extract enough information to uniquely identify an individual,  providing an immune fingerprint. The ``Immprint'' classifier analyzes this immune fingerprint to decide whether two samples were sampled from the same individual.}

\section*{Results}

Given two samples of peripheral blood containing respectively $M_1$ and $M_2$ T-cells, we want to distinguish between two hypothetical scenarios: either the two samples come from the same individual (``autologous'' scenario), or they were obtained from two different individuals (``heterologous'' scenario), see Fig.~\ref{fig:one}a.

TCR are formed by two protein chains $\alpha$ and $\beta$. They each present a region of high somatic variability, labeled CDR3$\alpha$ and CDR3$\beta$, randomly generated during the recombination process. These regions are coded by short sequences (around 50 nucleotides), which are captured by RepSeq experiments. The two chains are usually not sequenced together so that the pairing information between $\alpha$ and $\beta$ is lost. \thomas{Most experiments focus on the $\beta$ chain, and when not otherwise specified, the term ``receptor sequence'' in this paper will refer only to the nucleotide sequence of the TRB gene coding for this $\beta$ chain (which include CDR3$\beta$). Similarly, as most cells expressing the same beta chain are clonally related, we will be using the terms ``clone'' and ``clonotype'' to refer to set of cells with the same nucleotide TRB sequence, even if they were produced in separate generation events and are not real biological clones (since we have no means of distinguishing the two cases).} CDR3$\beta$ sequences are very diverse, with more than $10^{40}$ possible sequences \cite{Mora2016}. 
For comparison, the TCR$\beta$ repertoire of a given individual is composed of $10^8$ to $10^{10}$ unique clonotypes \cite{Qi2014,Lythe2016}. As a result, most of the sequences found in a repertoire are ``private''.

\subsection*{Immprint scores}

\begin{figure*}[!htbp]
  \centering
  \includegraphics[width=.8\linewidth]{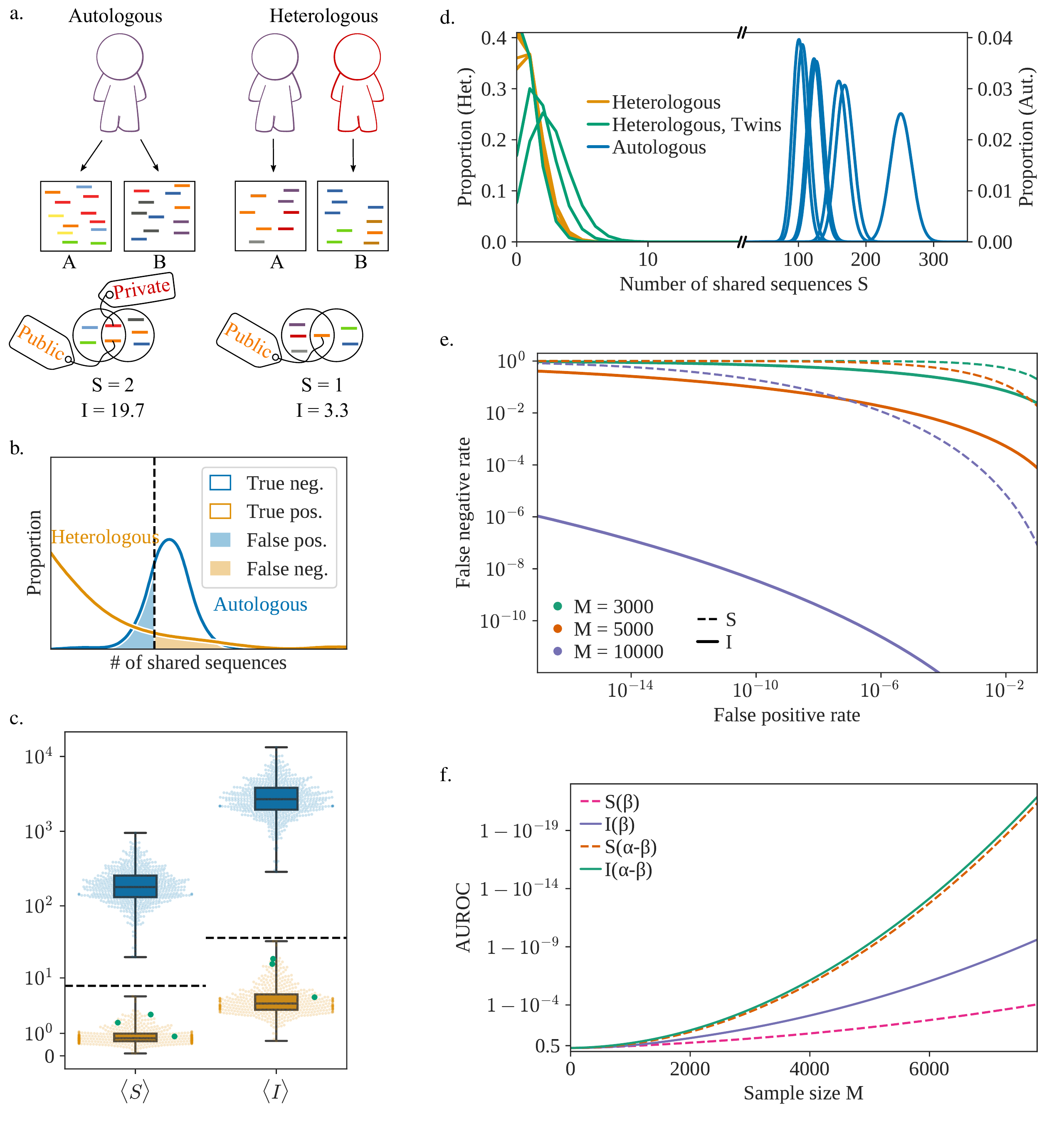}
  \caption{a) The two samples A and B can either originate from the same individual (autologous) or two different individuals (heterologous).  In both scenarios, sequences can be shared between the two samples, but their quantity and quality vary. b) \thomas{Schematic representation of the distribution of the $\mS$ or $\mI$ scores for multiple pairs of samples extracted from the same individual (in the autologous scenario) or the same pair of individuals (heterologous)}. The dashed vertical line represents the threshold value. c) Expected value of $\mS$ and $\mI$ for different pairs of samples, sampled from the same individual (in blue) or different ones (yellow). Green dots represent samples extracted from pairs of identical twins. The dashed lines represents the theoretical upper bound for heterologous repertoires (see Methods) for both $\mS$ and $\mI$ ($\gamma = 12$). d) \thomas{ Sampling distributions of $\mS$ for 6 different patients (autologous case, each blue curve is one patient) or 6 different pairs of patients (heterologous, each yellow or green curve is a pair of patients) for $M=5000$. The y-axis scale on the left is adapted to the heterologous distributions while the scale on the right corresponds to the (much wider) autologous ones. The 3 sampling distributions in green correspond to a pair of samples extracted from identical twins.} e) Detection Error Trade-off (DET) graph for both summary statistics and different sample sizes $M$. $\mI$ ($\gamma = 12$) outperforms $\mS$ in all scenarios.  f) AUROC (Area Under Receiver Operating Characteristic), as a function of $M$. The AUROC is a traditional measure of the quality of a binary classifier (a score closer to one indicates a better classifier). The results are shown for $\mS$ and $\mI$ both in the default case (only the $\beta$ chain considered) or for the full ($\alpha$-$\beta$) receptor.}
  \label{fig:one}
\end{figure*}

To discriminate between the autologous and heterologous scenarios, one can \thomas{simply} count the number of nucleotide receptor sequences shared between the two samples\thomas{, which we call $\mS$}. Samples coming from the same individual should have more receptors in common because T-cells are organized in clones of cells carrying the same TCR. By contrast, $\mS$ should be low in pairs of samples from different individuals, in which sharing is due to rare convergent recombinations. Appropriately setting a threshold to jointly minimize the rates of false positives and false negatives (Fig.~\ref{fig:one}b), we can use $\mS$ as a classifier to distinguish autologous from heterologous samples. \thomas{$\mS$ is not normalized for sequencing depth and values of $\mS$ should not be compared between samples of different  size.}

The $\mS$ score can be improved upon by exploiting the fact that some receptors are much more likely than others to be generated during V(D)J-recombination, with variations in generation probability ($P_\mathrm{gen}$, \cite{Murugan2012,Marcou2018,Sethna2020}) spanning 15 orders of magnitude. Public sequences (with high $P_\mathrm{gen}$) are likely to be found in multiple individuals \cite{venturi_2008}, while rare sequences (low $P_\mathrm{gen}$) are unlikely to be shared by different individuals, and thus provide strong evidence for the autologous scenario when found in both samples. To account for this information, we define the score:
\begin{equation}\label{Iscore}
\mathcal{I} = \sum_{\textrm{shared }s} \left[\ln\left(1/P_{\mathrm{gen}}(s)\right)-\gamma\right],
\end{equation}
which accounts for Shannon's ``surprise'' $\ln(1/P_{\rm gen})$---a measure of unexpectedness---associated with each shared sequence $s$, so that rare shared sequences count more than public ones. The constant $\gamma$ depends on the repertoire's clonal structure and is set to 12 in the following (see Methods for an information-theoretic derivation). $P_{\text{gen}}$ is computed using models previously trained on data from multiple individuals \cite{Marcou2018}. Small differences reported between the $P_{\text{gen}}$ of distinct individuals justify the use of a universal model \cite{Sethna2020}.

\subsection*{Measuring error rates}

We tested the classifiers based on the $\mS$ and $\mI$ scores on TCR$\beta$ RepSeq datasets from 656 individuals \cite{Emerson2017} \thomas{-- labeled according to their cytomegalovirus (CMV) serological status}. Sequences were downsampled to mimic experiments where $M_1=M_2=M$ cells were sampled and their receptors sequenced. \thomas{The frequency of a particular clonotype in the sample (the proportion of cells expressing a particular beta chain)  was estimated using the read counts of unique TCR$\beta$ sequences, and the mean values of $\mS$ and $\mI$ computed with a procedure designed to correct for the limited diversity of the sampled repertoire relative to the full repertoire, see Methods \ref{sec:methode_dataset_bias}}. Similar results may be obtained when $M_1$ and $M_2$ are different (see Methods). \thomas{The clones most often shared between two autologous samples are also the most clonally expanded -- and hence are probably antigen experienced. We verified that the sequence statistics of those expanded clonotypes did not differ from generic ones (Fig.~\ref{fig:rank_vs_pgen}).}

In Fig.~\ref{fig:one}c, we plot the mean value of $\mS$ (over many draws of $M=5000$ cells) for each individual (autologous scenario, in blue) and between pairs of different individuals (heterologous scenario, in yellow). The two scenarios are clearly discernable under both scores. 
This result holds for pairs of monozygotic twins obtained from a distinct dataset \cite{Pogorelyy2018} (green dots), consistent with previous reports that twins differ almost as much in their repertoires as unrelated individuals \cite{Zvyagin2014,Pogorelyy2018,Tanno2020}.
Heterologous scores (yellow dots) vary little, and may be bounded from above by a theoretical prediction (dashed line) based on a model of recombination \cite{elhanati_predicting_2018} (see Methods).
On the other hand, autologous scores (blue dots) show several orders of magnitude of variability across individuals. These variations stem from the clonal structure of the repertoire, and correlates with measures of diversity (Fig.~\ref{fig:comparison_diversity_measure}), which is known to vary a lot between individuals and correlates with age \cite{Britanova2016}, serological status, and infectious disease history \cite{sylwester_2005,khan_2002}.
To explore the worst case scenario of discriminability, hereafter we will focus on the individual with the lowest autologous $\mS$ found in the dataset.

The sampling process introduces an additional source of variability within each individual. Two samples of blood from the same individual do not contain the exact same receptors, and the values of $\mS$ and $\mI$ is expected to vary between replicates. \thomas{Examples of these variations are shown in Fig.~\ref{fig:one}d. The blue (respectively yellow) curves correspond to the sample distributions in the autologous (heterologous) scenario for different individuals (pairs of individuals).} The distribution of $\mS$ is well-approximated by a Poisson distribution, while $\mI$ follows approximately a compound distribution of a normal and Poisson distributions (see Methods for details). 
Armed with these statistical models of variations, we can predict upper bounds for the false negative and false positive rates.
As seen from the detection error trade-off (DET) graph Fig.~\ref{fig:one}e,  the Immprint classifier performs very well for a few thousand receptors with an advantage for $\mI$.

With $10,000$ cells, corresponding to $\sim 10$ $\mu$L of blood, Immprint may simultaneously achieve a false positive rate of $< 10^{-16}$ and false negative rate of $< 10^{-6}$, allowing for the near-certain identification of an individual  \thomas{based on the $\mI$ score} in pairwise comparisons against the world population $\sim 10^{10}$. When a large reference repertoire has been collected ($M_1=1,000,000$, corresponding to $\sim 1$mL of blood), an individual can be identified with just 100 cells (Fig.~\ref{fig:det_full_dataset}).

The AUROC estimator (Area Under the Curve of the Receiver Operating Characteristic), a typical measure of a binary classifier performance, can be used to score the quality of the classifier with a number between 0.5 (chance) and 1 (perfect classification). 
The $\mI$ score outperforms the $\mS$ score (Fig.~\ref{fig:one}f), particularly above moderate sample sizes ($M\approx 5000$).
Both scores can be readily generalized to the case of paired receptors TCR$\alpha\beta$, when the pairing of the two chains is available (through single-cell sequencing \cite{dash_2011,Redmond2016,Grigaityte2017} or computational pairing \cite{Howie2015}), using $P_{\mathrm{gen}}\left(\alpha, \beta \right) = P_{\mathrm{gen}}\left(\alpha\right) \times P_{\mathrm{gen}}\left(\beta\right)$ \cite{Dupic2019} for the generation probability of the full TCR. Because coincidental sharing of both chains is substantially rarer than with the $\beta$ chain alone, using the paired chain information greatly improves the classifier.

\thomas{While this paper focuses on T-cells and TCR sequences, the structure of the B-cell receptors (BCR) repertoire is very similar to the TCR repertoire and we expect to find qualitatively similar results. As an example we use the dataset obtained in Ref.~\cite{Briney2019} to measure $\mS$ and $\mI$ for IGH chains (forming half of the BCR receptor), Fig~\ref{fig:BCR_Immprint}. We see that 5000 IgG+ memory B-cells are enough to reliably identify the individuals in the study. However, B-cells are 5 times less common in peripheral blood than T-cells, and somatic hypermutation tends to distort the statistics of receptors, reducing the reliability of our classifier. Hence, for practical applications, T-cells are a better means of identification.}

\begin{figure*}[!htb]
  \centering
  \includegraphics[width=.8\linewidth]{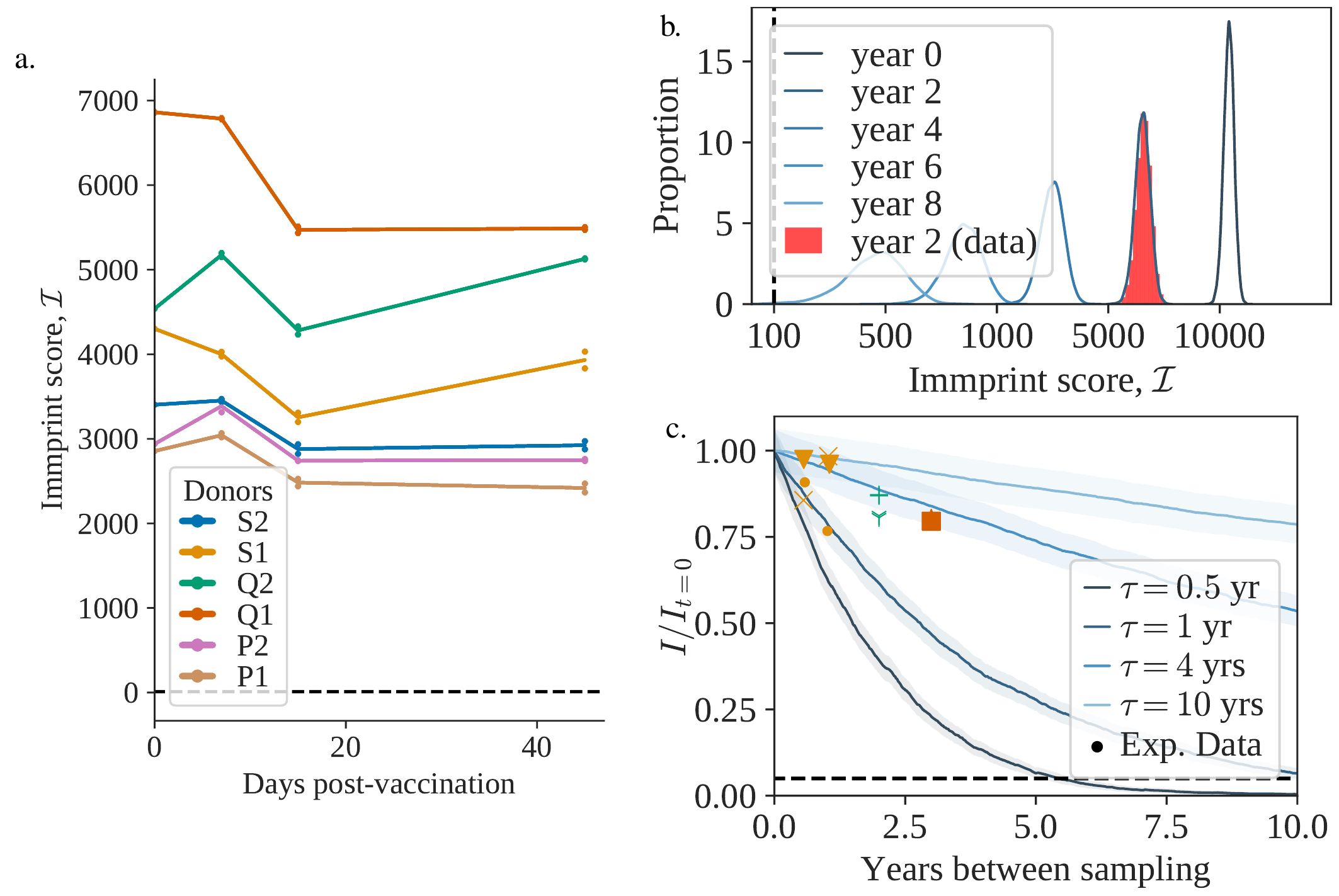}
  \caption{a) Evolution of $\mI$ (M=5000) during vaccination, between a sample taken at day $0$ (vaccination date) and at a later timepoint. Each color represents a different individual. Each pair timepoint/individual has two biological replicates. The dashed line represents the threshold value. b) Evolution of $\mI$ between a sample taken at year 0 and a later timepoint. \thomas{The red histogram corresponds to one of the individuals sampled in \cite{Pogorelyy2018} and the blue curves show theoretical estimates, fitted to match ($\tau = 0.66$).} c) Evolution of the (normalized) mean of $\mI$ (M=5000) as a function of time for different values of the turnover rate $\tau$. The dashed line represents the threshold value divided by the smallest value of $\mI_{t=0}$ (M=5000) in the data. The data points were obtained from the datasets \cite{Chu2019} (yellow), \cite{Pogorelyy2018} (green) and \cite{Britanova2016} (orange). Different markers indicate different individuals.}
  \label{fig:three}
\end{figure*}

\subsection*{Evolution with time}

The previous results used samples obtained at the same time. However, immune repertoires are not static: interaction with pathogens and natural aging modify their composition. The evolution of clonal frequencies will decrease Immprint's reliability with time, especially if the individual has experienced immune challenges in the meantime.

To study the effect of short-term infections, we analyzed an experiment where 6 individuals were vaccinated with the yellow fever vaccine, which is regarded as a good model of acute infection, and their immune system was monitored regularly through blood draws \cite{Pogorelyy2018}. We observe an only moderate drop in $\mI$ caused by vaccination (Fig.~\ref{fig:three}~a). 
This is consistent with the fact that infections lead to the strong expansion of only a limited number of clones, while the rest of the immune system stays stable \cite{DeWitt2015, Wolf2018, Qi2016, Sycheva2018}.
While other types of infections, auto-immune diseases, and cancers may affect Immprint in more substantial ways, our result suggests that it is relatively robust to changes in condition.

\medskip

We then asked how stable Immprint is over long times.
Addressing this issue is hampered by the lack of longitudinal datasets over long periods, so we turn to mathematical models \cite{Borghans2007,Thomas-Vaslin2008,Desponds2016, Lythe2016,DeGreef2020} to describe the dynamics of the repertoire. Following the model of fluctuating growth rate described in Ref.~\cite{Desponds2016}, we define two typical evolutionary timescales for the immune system: $\tau$, the typical turnover rate of T-cell clones, and $\theta$, which represents the typical time for a clonotype to grow or shrink by a factor two as its growth rate fluctuates. The model predicts a power-law distribution for the clone-size distribution, with exponent $-1-\tau/2\theta$. This exponent has been experimentally measured to be $\approx -2$ \cite{Mora2016}, which leaves us with a single parameter $\tau$, and $\theta=\tau/2$.  An example of simulated evolution of Immprint with time is shown in Fig.~\ref{fig:three}~b. The highlighted histogram represents a data point at two years obtained from \cite{Pogorelyy2018}. \thomas{While a fit is possible for this specific individual ($\tau = 0.66$ in Fig.~\ref{fig:three}~b), the $\tau$ parameter is not universal, and we expect it to vary between individuals, especially as a function of age. In Fig.~\ref{fig:three}c we explore the effect on the stability of Immprint for a range of reasonable values for the clone turn-over rate $\tau$, from 6 months to 10 years, encompassing  both previous estimates of the parameter~\cite{Desponds2016} and measured turnover rates for different types of T-cells \cite{DeBoer2013a}. While Fig.~\ref{fig:three} focuses on $\mI$, the behaviour is similar for $\mS$ (see Fig.~\ref{fig:fig2_but_with_S}). We observe that under this model, for most individuals and bar exceptional events, Immprint should conserve its accuracy for years or even decades.} 

\section*{Discussion}

\thomas{In summary, the T-cells present in small blood samples provide a somatic and long-lived barcode of human individuality, which is robust to immune challenges and stable over time. While the uniqueness of the repertoire was a well known fact, we demonstrated that the most common T-cells clones are still diverse enough to uniquely define an individual and frequent enough to be reliably sampled multiple times.} Unlike genome sequencing, repertoire sequencing can discriminate monozygotic twins with the same accuracy as unrelated individuals. However, a person's unique immune fingerprint can be completely wiped out by a hematopoietic stem cell transplant~\cite{buhler_2020}. \thomas{The different datasets used cover a range of different sequencing methods \ref{seq:datasets}, however different approaches may lead to slightly different threshold choices. In particular, in practical implementations, sequencing depth is an important concern. One needs enough coverage to sequence TCR$\beta$ genes from as many as possible of the T-cells present in the sample, in order to measure a more precise a immune fingerprint. In addition, the specific calculations presented here only apply to peripheral blood cells. Specific cell types or cells extracted from tissue samples may have different clonal distributions and potentially different receptor statistics. For example the value of $\mS$ in the autologous case varies between CD4+ and CD8+ T-cells (Fig.~\ref{fig:CD4_CD8}), different individuals remain distinguishable using each subset.}

Immprint is implemented in a python package and webapp (see Methods) allowing the user to determine the autologous or heterologous origin of a pair of repertoires. Beyond identifying individuals, the tool could be used to check for contamination or labelling errors between samples containing TCR information.
The repertoire information used by Immprint can be garnered not only from RepSeq experiments, but also from RNA-Seq experiments, which contain thousands of immune receptor transcripts \cite{li_2017,Bolotin2017b}. 
Relatively small samples of immune repertoires are enough to uniquely identify an individual even among twins, with potential forensics applications. At the same time, unlike genetic data from genomic or mRNA sequencing, Immprint provides no information about kin relationships, very much like classical fingerprints, and avoids privacy concerns about disclosing genetic information shared with non consenting relatives.

\section*{Acknowledgments}
The study was supported by the European Research Council COG 724208. A.A.M and M.V.P. are supported by RSF-20-15-00351. The authors are grateful for the help of Natanael Spizak with the analysis of BCR repertoire datasets.

\bibliographystyle{pnas}

\onecolumngrid

 % methods
\newpage
\section*{Methods}

\subsection*{Datasets \& Pre-processing}
\label{seq:datasets}
We use five independant RepSeq datasets in this study:
{\em (i)} genomic DNA from Peripheral blood mononuclear cells (PBMCs) from 656 healthy donors 
  \cite{Emerson2017}; 
  {\em (ii)} cDNA of PBMCs sampled from three pairs of twins, before and after a yellow-fever
  vaccination \cite{Pogorelyy2018}; 
  {\em (iii), (iv)} two longitudinal studies of healthy adults
  \cite{Chu2019, Britanova2016};\thomas{{\em (v)} cDNA dataset of IGH genes (B-cells) from 9 individuals, with multiple replicates \cite{Briney2019}.} 

  CDR3 nucleotide sequences were extracted with MIGEC \cite{Shugay2014} (for the second dataset) coupled with MiXCR \cite{Bolotin2015}. We also extract the frequency of reads from the three datasets. The non-productive sequences were discarded (out-of-frame, non-functional V gene, or presence of a stop codon). The generation probability ($P_{\mathrm{gen}}$) was computed using OLGA \cite{Sethna2019}, with the
default TCR$\beta$ model. \thomas{The frequency of each clone was estimated by summing the frequencies of all reads that shared the same nucleotide CDR3 sequence and identical V,J genes.}

The preprocessing code is distributed on the Git repository associated with the paper. We also developed a command-line tool (\url{https://github.com/statbiophys/immprint}) that discriminates between sample origins, and a companion webapp (\url{https://immprint.herokuapp.com}).

\subsection*{Discrimination scores}

\label{sec:methods_summary_statistics}
To discriminate between the autologous and heterologous scenarios, we introduce a log-likelihood ratio test between the two possibilities:
\beq
\mathcal{I}= \sum_s \ln\frac{P(y_1(s),y_2(s)|{\rm autologous})}{P(y_1(s),y_2(s)|{\rm heterologous})},
\eeq
where $y_1(s)=1$ if the sequence $s$ is found in sample 1, and 0 otherwise; likewise $y_2(s)=1$ if $s$ is in sample 2.
The sum runs over all potential sequences $s$, including unseen ones.
To be present in a sample, a sequence $s$ first has to be present in the repertoire. This occurs with probability $1-(1-p(s))^{N_{\rm c}}$, where $N_{\rm c}$ is the total number of clonotypes in the repertoire, and $p(s)$ is the probability of occurence of sequence $s$ (resulting from generation and selection, see below). Second, it must be picked in a sample of size $M$, with probability $1-(1-f)^{M}\approx Mf$ (assuming $Mf\ll 1$) depending on its frequency $f$, which is distributed according to the clone size distribution $\rho(f)$. We checked that $f(s)$ and $P_{\rm gen}(s)$ were not correlated (Fig.~\ref{fig:rank_vs_pgen}) \thomas{and that the effects of a shared infection between different individuals were limited to a handful of clones (Fig.~\ref{fig:rank_vs_pgen_plusplus})}. 
Then one can write
\begin{align}
P(y_1(s)=1,y_2(s)=1\ |\ {\rm autologous})&
\approx \left(1-e^{-N_{\rm c}p(s)}\right)M_1M_2\int df\,\rho(f)\,f^2,\\
P(y_1(s)=1,y_2(s)=0\ |\ {\rm autologous})&
\approx \left(1-e^{-N_{\rm c}p(s)}\right)\frac{M_1}{N_{\rm c}}\textrm{ and }1\leftrightarrow 2,\\
P(y_1(s)=0,y_2(s)=0\ |\ {\rm autologous})&
\approx 1-\left(1-e^{-N_{\rm c}p(s)}\right)\frac{M_1+M_2}{N_{\rm c}},
\end{align}
where we've used $\int df\,\rho(f)\,f = 1/N_{\rm c}$.
For the heterologous case the probability factorizes as:
\beq
P(y_1(s),y_2(s)\ |\ {\rm heterologous})=P_1(y_1(s))P_2(y_2(s)),
\eeq
with
\beq
P_a(y_a(s)=1)\approx \left(1-e^{-N_{\rm c}p(s)}\right) \frac{M_a}{N_{\rm c}},\quad a=1,2.
\eeq
Since only the term $y_1(s)=y_2(s)=1$ (shared sequences) is different between the autologous and heterologous cases, we obtain:
\beq
\mathcal{I}= \sum_{\textrm{shared } s} \left[ \ln(N_c^2\<f^2\>) - \ln \left(1-e^{-N_{\rm c}p(s)}\right)\right].
\eeq
Further assuming $N_cp(s)\ll 1$, and $p(s) = P_{\rm gen}(s)q^{-1}$ (where $q$ accounts for selection \cite{elhanati_predicting_2018} and $P_{\rm gen}(s)$ is the probability of sequence generation \cite{Marcou2018}), the score simplifies to Eq.~\ref{Iscore}, with $\gamma=-\ln(qN_c\<f^2\>)=\ln(q^{-1}\<f\>/\<f^2\>)$. The factor $\gamma$ depends on unknown parameters of the model, but can be estimated assuming a power-law for the clone size distribution \cite{Touzel2020}, $\rho(f)\propto f^{-2}$ extending from $f=10^{-11}$ to $f=0.01$, and $q=0.01$ \cite{elhanati_predicting_2018}, yielding $\gamma\approx 12.24$. Alternatively we optimized $\gamma$ to minimize the AUROC, yielding $\gamma\approx 15$ (SI Fig.~\ref{fig:Igamma}). Since performance degrades quickly for larger values, we conservatively set $\gamma=12$.

\subsection*{Estimating mean scores from RepSeq datasets}
\label{sec:methode_dataset_bias}

\thomas{The sampling of $M$ cells from blood is simulated using large repertoire datasets. In a bulk repertoire sequencing dataset, the absolute number of cells for each clonotype  (cells with a specific receptor) is unknown, but the fraction of each clonotype can be estimated using the proportion of reads that are associated with this specific receptor.} To estimate the autologous $\mS$ and $\mI$ of two samples of size $M_1$ and $M_2$ in the absence of true replicates, we computed their expected values from a single dataset containing $N$ reads, from which two random subsamples of sizes $M_1$ and $M_2$ were taken.
The mean value of $\mS$ is equal to $\langle \mS \rangle = \sum_s (1 - (1 - f(s))^{M_1}) (1 - (1 - f(s))^{M_2})$, where $f(s)$ is the true (and unknown) frequency of sequence $s$. A naive estimate of $\<mS\>$ may be obtained by repeatly resampling subsets of sizes $M_1$ and $M_2$ from the observed repertoire, calculate $\mS$ for each draw, and average. One get the same result by replacing $f(s)$ by $\hat{f}_s=n(s)/N$ in the previous formula, where $n(s)$ is the number of $s$ reads in the full dataset, and $N=\sum_s n(s)$. However, this naive estimate leads to a systematic overestimate of the sharing (visible when compared with biological replicates, see Fig.~\ref{fig:S_unbiased_naive}), simply because this procedure overestimates the probability of resampling rare sequences, in particular singletons whose true frequency may be much lower that $1/N$.
A similar bias occurs when computing $\mI$.
To correct for this bias, we look for a function $h(n)$
that satisfies for all $f$:
\beq\label{cond1}
\<h(n)\>\equiv \sum_n {N\choose n} f^n (1-f)^{N-n}h(n) = (1 - (1 - f)^{M_1})\ (1 - (1 - f)^{M_2}),
\eeq
so that $\<\mS\>$ and $\<\mI\>$ can be well approximated by:
\begin{align}
  \<\mS\>&\approx \sum_s h(n(s)),\label{eq:ma}\\
  \<\mI\>&\approx -\sum_s h(n(s)) \left[\ln(1/P_{\rm gen}(s))-\gamma\right].
\end{align}

Expanding the right-hand side of Eq.~\ref{cond1} into 4 terms, we find that $h(n)=1-g_{M_1}(n)-g_{M_2}(n)+g_{M_1+M_2}(n)$ satisfies Eq.~\ref{cond1} provided that:
\beq
\sum_n {N \choose n} f^n (1 - f)^{N - n} g_M(n) = (1 - f)^M.
\eeq
Under the change of variable $x = f/(1-f)$, the expression becomes:
\beq
\sum_n {N \choose n} x^n g_M(n) = (1 + x)^{N - M} = \sum_n {N - M \choose n} x^n.
\eeq
Identifying the polynomial coefficients in $x^n$ on both sides yields:
\beq
g_M(n) = \left. {N - M \choose n} \middle/ {N \choose n} \right. .
\eeq
These corrected estimates agree with the direct estimates using biological replicates (Fig.~\ref{fig:S_unbiased_naive}).

Similarly, $\<\mS\>$ and $\<\mI\>$ in heterologous samples can be estimated using:
\begin{align}
  \label{eq:estimate_S_I}
  \<\mS\>&\approx \sum_s [1-g_{M_1}(n(s))][1-g_{M_2}(n'(s))],\\
  \<\mI\>&\approx \sum_s [1-g_{M_1}(n(s))][1-g_{M_2}(n'(s))]\left[\ln(1/P_{\rm gen}(s))-\gamma\right].
\end{align}
where $n(s)$ and $n'(s)$ are the empirical counts of sequence $s$ in the two samples.

\subsection*{Theoretical upper bound on heterologous scores}
\label{sec:method_predicting_sharing}
When the two samples were extracted from two different individuals (heterologous scenario), we can use the universality of the recombination process to give upper bounds on the values of $\mS$ and $\mI$. These bounds are represented by the dashed lines in Fig\ref{fig:one}c). If two samples of respectively $M_1$ and $M_2$ cells, containing $T_1 \leq M_1$ and $T_2 \leq M_2$ unique sequences are extracted from two different individuals, the number of shared sequences between them is given by \cite{elhanati_predicting_2018}:
\begin{align}\label{eq:mh}
  \<\mathcal{S}\>_{\rm heterologous} &\leq \sum_s \left(1 - \left(1 - p(s)\right)^{T_1}\right)\left(1 - \left(1 - p(s)\right)^{T_2}\right)\\ &\lessapprox T_1 T_2 \sum_s p(s)^2=T_1T_2 \<p(s)\> \leq M_1 M_2 \<p(s)\>.
\end{align}
$p(s)$ is the probability of finding a sequence $s$ in the blood. Following \cite{elhanati_predicting_2018}, we make the approximation $p(s) = P_{\mathrm{gen}}(s) q^{-1}$, where the $q=0.01$ factor is the probability that a generated sequences passes selection. Then $\<p(s)\>$ can be estimated from the mean over generated sequences.
Similarly, we can estimate $\mI$ as
\begin{align}
 \<\mathcal{I}\>_{\rm heterologous}  &\lessapprox T_1 T_2 \sum_s p(s)^2 \left[\ln\left(1/P_{\mathrm{gen}}(s)\right) -\gamma\right] \\ &=-T_1T_2 \<p(s)[\gamma+\ln(qp(s))]\> \leq -M_1M_2 \<p(s)[\gamma+\ln(qp(s))]\>,
\end{align}
which is also estimated from the mean over generated sequences.

\subsection*{Error rate estimates}

To make the quantitative predictions shown in Fig.~\ref{fig:one}, we need to constrain the tail behavior of the distributions of $\mS$ and $\mI$, for the two scenarios.

The $\mS$ statistic can be rewritten as a sum of Bernouilli variables over all possible sequences, each with a parameter corresponding to its probability of being present in both samples, either in the autologous or the heterologous case. Therefore $\mS$ is a Poisson binomial distribution, a sum of independent Bernouilli variables with potentially different parameters. The variance and tails of that distribution are bounded by those of the Poisson distribution with the same mean, denoted by $m_a$ for the autologous case, and $m_h$ for the heterologous case (Fig.~\ref{fig:poisson_approx}).

Thanks to that inequality, the rates of false negatives and false positives for a given threshold $r$ are bounded by:
\beq
P(\mathcal{S} < r | \text{autologous}) \leq Q(r + 1, m_{a}), \qquad P(\mathcal{S} > r | \text{heterologous}) \leq 1 - Q(r + 1, m_{h}),
\eeq
where
$Q$ is the regularized gamma function, which appears in the cumulative distribution function of the Poisson distribution.
The mean autologous score $m_a$ is estimated from experimental data: we use the smallest value of $\langle \mathcal{S} \rangle$ in the Emerson dataset and Eq.~\ref{eq:ma}. To compute $m_h$, we use the semi-theoretical prediction made using the universality of the recombination process, Eq.~\ref{eq:mh}.

Similarly, $\mI$ can be viewed as a sum of $\mS$ independent random variables, all following the distribution of $\ln(1/P_\mathrm{gen})-\gamma$. However, this distribution differs in the two scenarios. Sequences shared between more than one donor have an higher probability of being generated, their $\ln(P_\text{gen})$ distribution has higher mean and smaller variance (Fig.~\ref{fig:pgen_distribution}).

The sum is composed of a relatively large number of variables in most realistic scenarios. Hence, we rely on the central limit theorem to approximate it by a normal distribution, of mean and variance proportional to $\mS$. Explicitly:

\beq
P(\mI < r | \text{autologous}) = \frac{1}{2} \sum_{\mS = 0}^\infty \frac{\left(m_a\right)^\mS e^{-m_a}}{\mS!} \left( 1 + \mathrm{erf}\!\left(\frac{r - \mS\langle \ln(1/P_\text{gen})-\gamma  \rangle}{\sqrt{2 \mS \mathrm{Var}\!\left[\ln(1/P_\text{gen})-\gamma \right]}} \right) \right),
\eeq
\beq
P(\mI > r | \text{heterologous}) = \frac{1}{2} \sum_{\mS = 0}^\infty \frac{\left(m_h\right)^\mS e^{-m_h}}{\mS!} \left( 1 - \mathrm{erf}\!\left(\frac{r - \mS\langle \ln(1/P_\text{gen})-\gamma \rangle_\text{shared}}{\sqrt{2 S \mathrm{Var}\!\left[ \ln(1/P_\text{gen})-\gamma \right]_\text{shared}}} \right) \right).
\eeq
The AUROC are computed based on these estimates, by numerically integrating the true positive rate $P(\mS,\mI < r | \text{heterologous})$ with respect to the false negative rate $P(\mS,\mI < r | \text{autologous})$ as the threshold $r$ is varied.

\subsection*{Modeling the evolution of autologous scores}
We use the model of Ref.~\cite{Desponds2016} to describe the dynamics of individual T-cell clone frequencies $f$ under a fluctuating growth rate reflecting the changing state of the environment and the random nature of immune stimuli:
\begin{equation}\label{eq:dyn}
 \frac{df}{dt} = \left[- \frac{1}{\tau} + \frac{1}{2\theta} + \frac{1}{\sqrt{\theta}}{\eta(t)}\right] f(t),
\end{equation}
where $\eta(t)$ is a Gaussian white noise with $\<\eta(t)\>=0$ and $\<\eta(t)\eta(t')\>=\delta(t-t')$.

With the change of variable $x=\ln(f)$, these dynamics simplify to a simple Brownian motion in log-frequency: $\partial_t x= -\tau^{-1}+\theta^{-1/2}\eta(t)$. In that equation, $\tau$ appears as the decay rate of the frequency, while $\theta$ is the timescale of the noise, interpreted as the typical time it takes for the frequency to rise or fall by a logarithmic unit owing to fluctuations. Considering a large population of clone, each with their independent frequency evolving according to Eq.~\ref{eq:dyn}, and a source term at small $f$ corresponding to thymic exports, one can show that the steady-state probability density function of $f$ follows a power-law \cite{Desponds2016}, $\rho(f)\propto f^{-\alpha}$, with exponent $\alpha=1+2\theta/\tau$. $\alpha$ was empirically found to be $\approx 2$ in a wide variety of immune repertoires \cite{Weinstein2009,Mora2016,Oakes2017,Touzel2020}, implying $2\theta\approx \tau$. The turn-over time $\tau$ is unknown, and was varied from $1/2$ year to 10 years in the simulations.

We simulated the evolution of human TRB repertoires by starting with the empirical values of the frequencies of each observed clones, $f(s,0)=\hat f(s,0)=n(s,0)/N$ from the analysed datasets. A sample of size $M$ was randomly selected with respect to these frequencies, and the frequencies of the clones captured in that sample were then evolved with a time-step of 2 days using Euler-Maruyama's method, which is exact in the case of Brownian motion. Clones with frequencies falling below $10^{-11}$ (corresponding to a single cell in the organism) were removed. At each time $t>0$, we measured the mean value of $\mathcal{S}$ with the formula $\sum_s (1 - (1 - f(s, t))^{M})$ where the sum runs over the sequences captured in the initial sample.

 % sifigs
\setcounter{table}{0}
\renewcommand{\thetable}{S\arabic{table}}%
\setcounter{figure}{0}
\renewcommand{\thefigure}{S\arabic{figure}}%

\begin{figure*}[!htbp]
\begin{center}
\includegraphics[width=1.\linewidth]{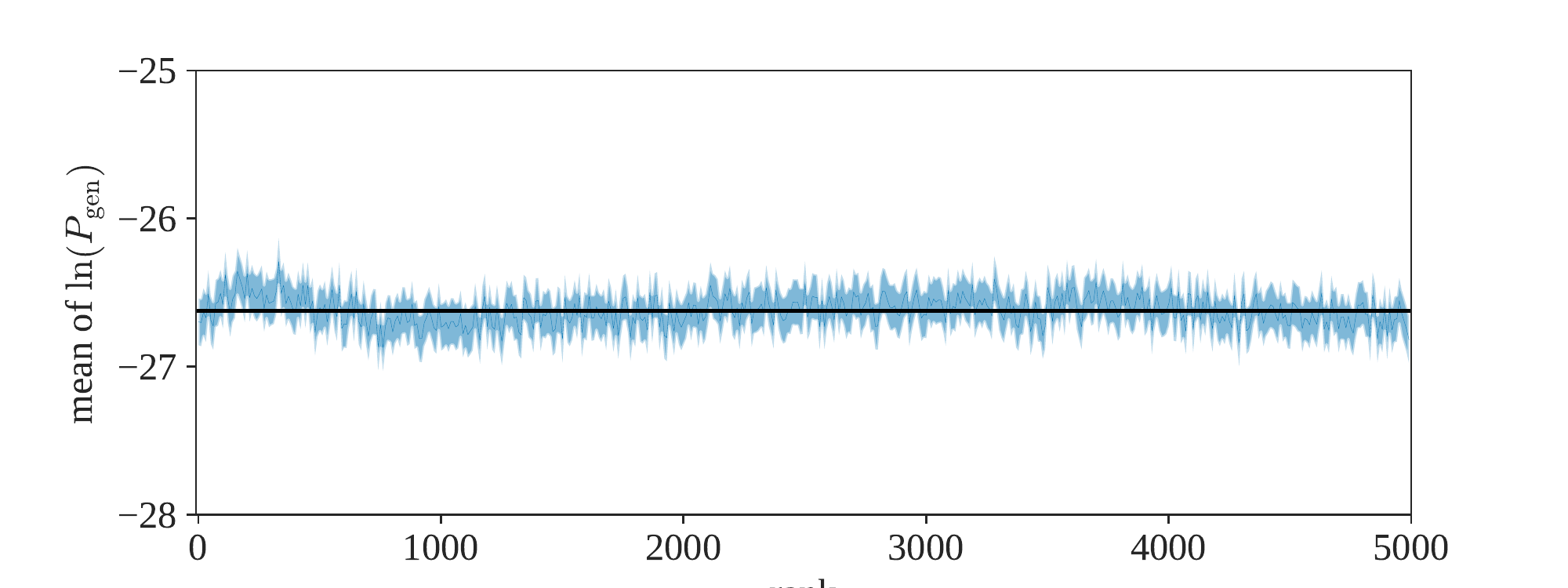}
\caption{\thomas{Mean value of $\log P_\mathrm{gen}$ as a function of the
  rank of the clonotype (from most abundant to least abundant) from
  \cite{Emerson2017}. The black line represents the mean of $\log P_\mathrm{gen}$ for naive clones. The statistic for the top-clones (low rank) is similar to the one for the naive clones.}
  \label{fig:rank_vs_pgen}
}
\end{center}
\end{figure*}

\begin{figure*}[!htbp]
\begin{center}
\includegraphics[width=.8\linewidth]{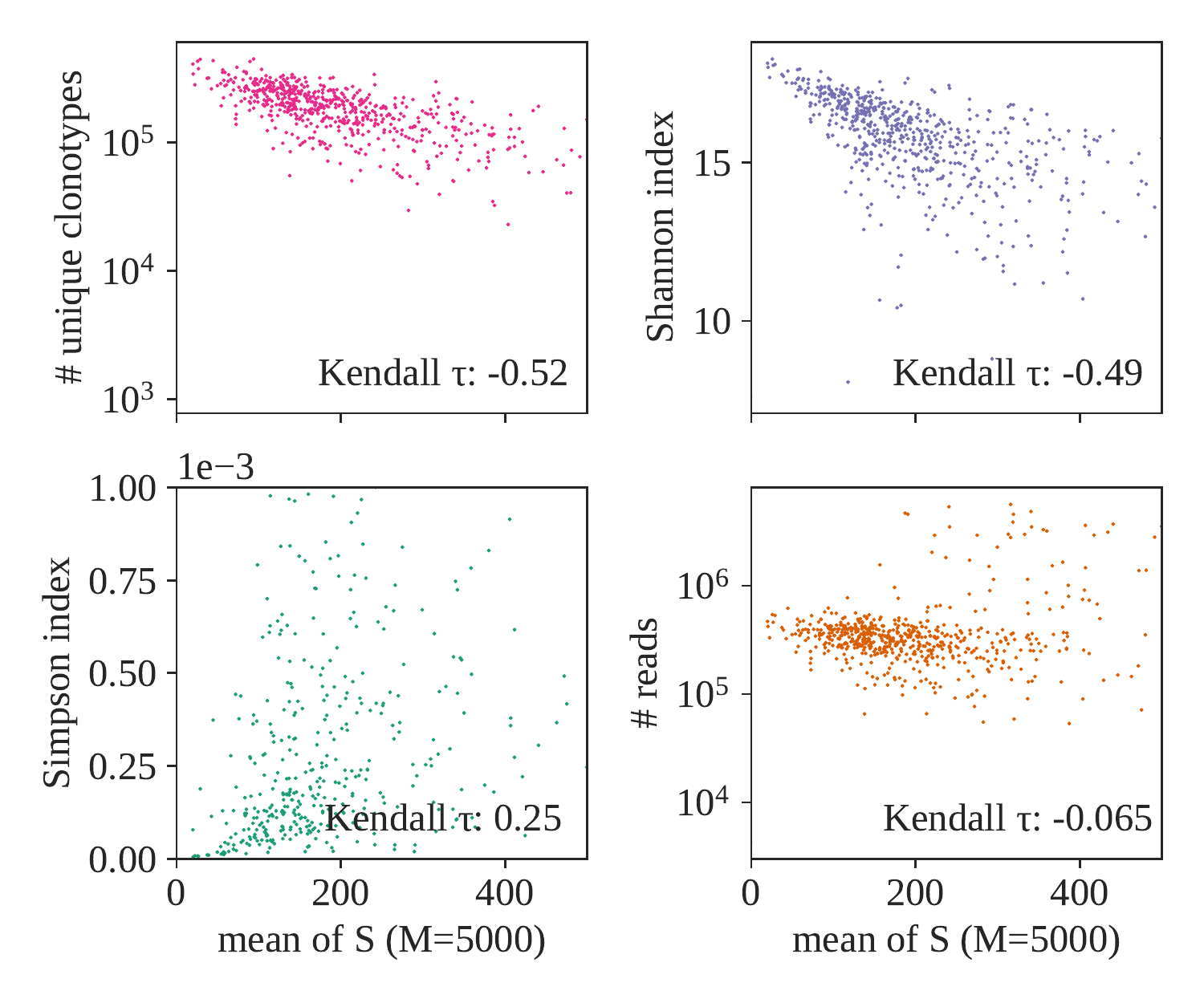}
\caption{Comparison between the mean of $\mS$ (autologous case), and three
  common diversity measures: the number of unique sequences found in
  the dataset (top left), the Shannon index, $-\sum \hat f_s \ln \hat f_s$ (top
  right), the Simpson index (bottom left), and the total number of
  reads in each datasets (bottom right). All the diversity measures
  show a strong correlation with $\mS$, but the correlation with the
  sequencing depth is low.
\label{fig:comparison_diversity_measure}
}
\end{center}
\end{figure*}

\begin{figure*}[!htbp]
\begin{center}
\includegraphics[width=.8\linewidth]{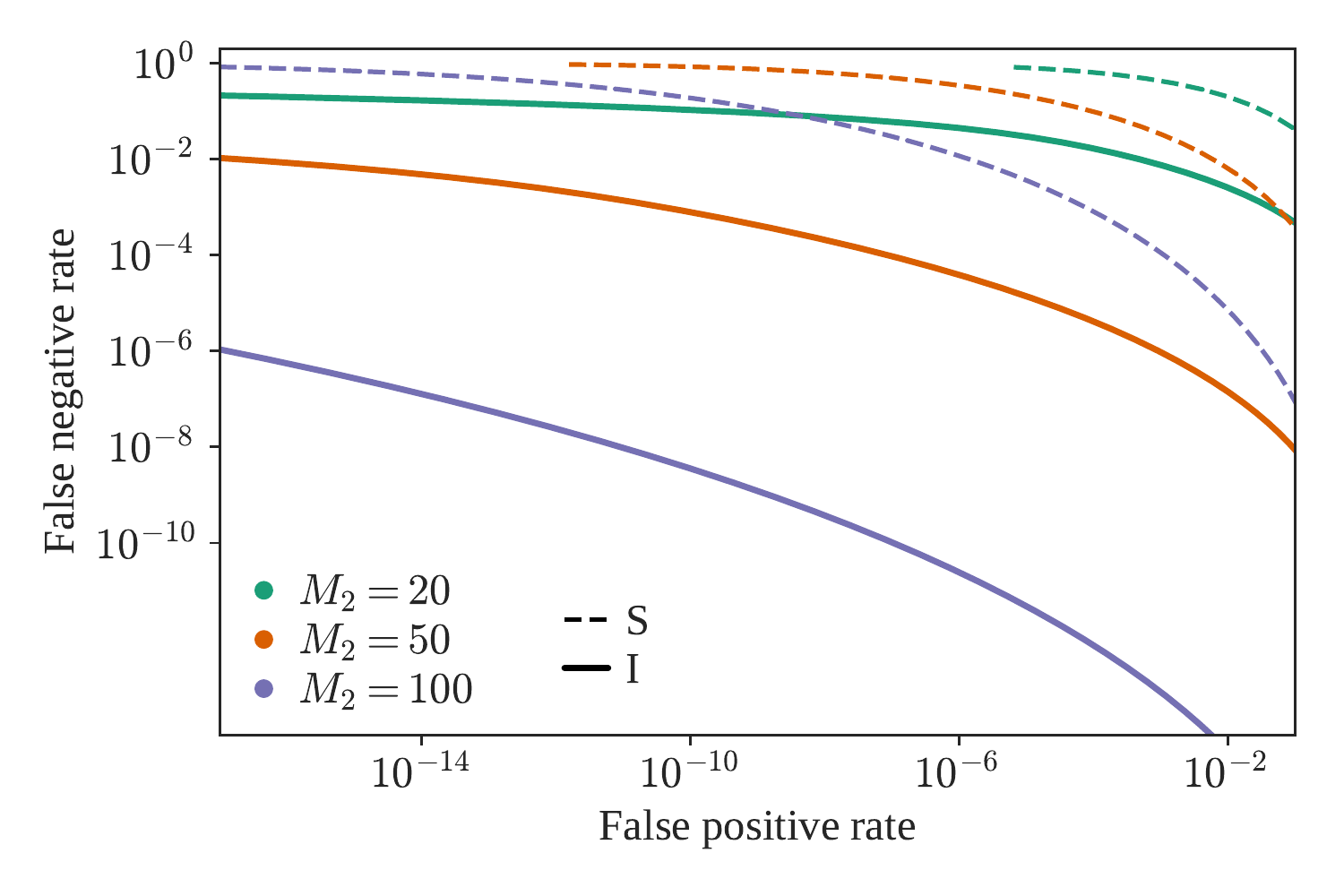}
\caption{Detection Error Trade-off (DET) graph for both summary
  statistics, between a large sample (full dataset, $M_1 = 10^6$) and
  a smaller one, of size $M_2=M$.
\label{fig:det_full_dataset}
}
\end{center}
\end{figure*}

\begin{figure*}[!htpb]
    \begin{center}
      \includegraphics[width=.8\linewidth]{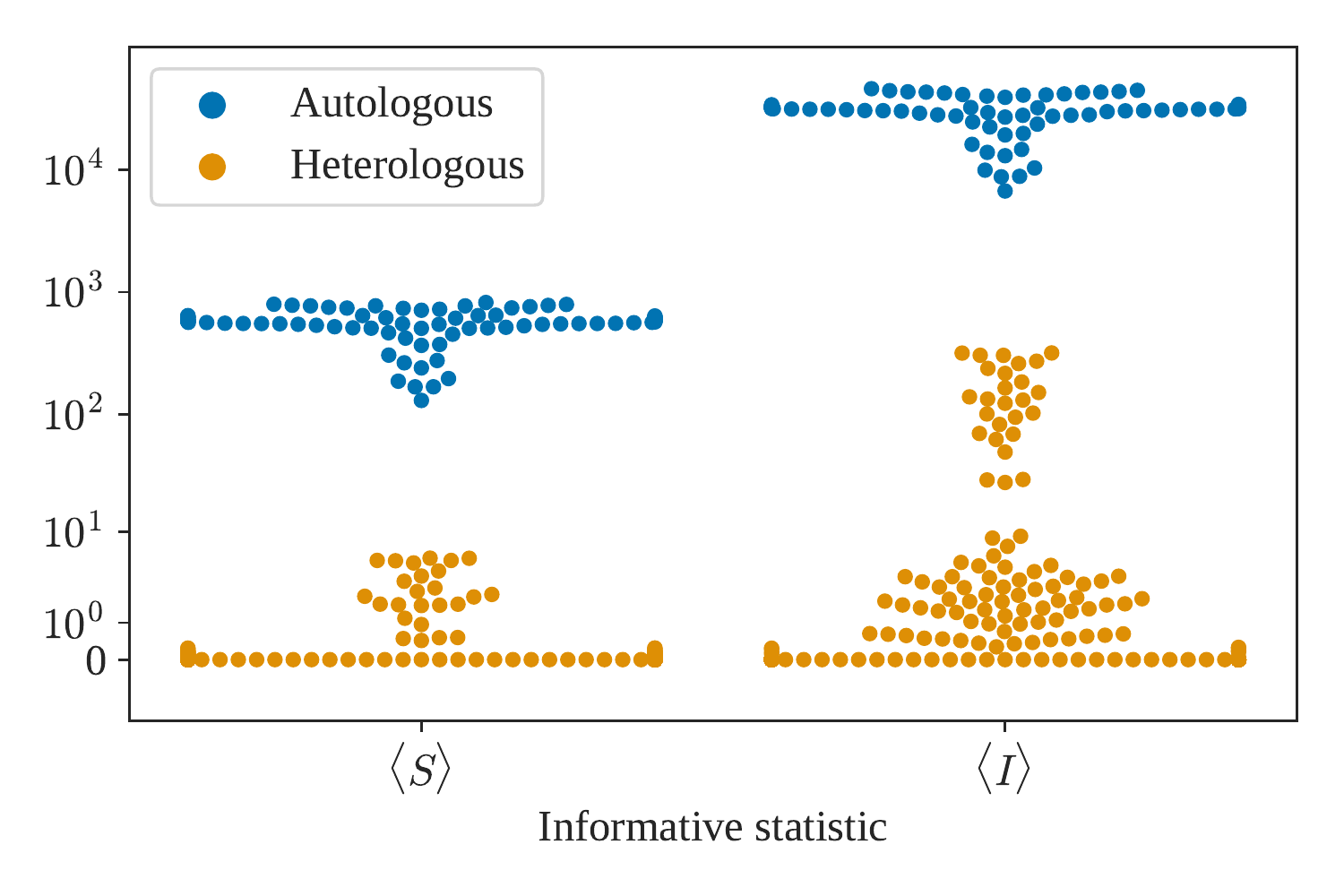}
      \caption{\thomas{Mean value of $\mS$ (left) and $\mI$ (right)
          for the IGH chain, part of the B-cell receptor, for sample
          sizes $M=5000$. The IGH sequences used are restricted to
          IgG+ B-cells (selected according to their CH gene). The
          sequences were obtained from 8 different individuals (6
          biological replicates each) in the dataset from \cite{Briney2019}. Autologous (blue) and heterologous (yellow) are well separated.}}
      \label{fig:BCR_Immprint}
    \end{center}
  \end{figure*}

\begin{figure*}[!htb]
  \centering
  \includegraphics[width=1. \linewidth]{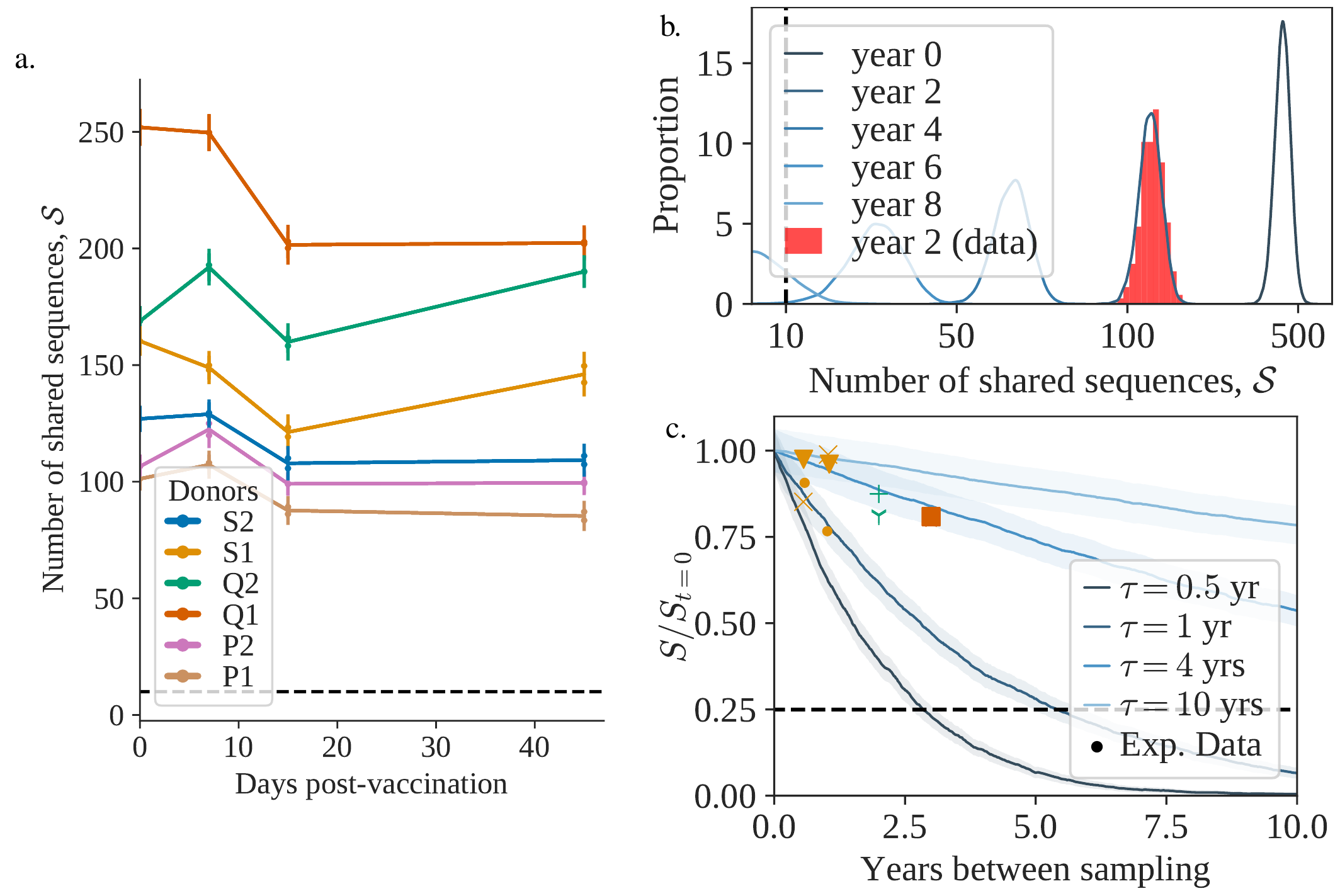}
  \caption{\thomas{a) Evolution of $\mS$ ($M=5000$) during
      vaccination, between a sample taken at day $0$ (vaccination
      date) and at a later timepoint. Each color represents a different individual. Each pair timepoint/individual has two biological replicates. The dashed line represents the threshold value. b) Evolution of $\mS$ between a sample taken at year 0 and a later timepoint. \thomas{The red histogram corresponds to one of the individuals sampled in \cite{Pogorelyy2018} and the blue curves show theoretical estimates, fitted to match ($\tau = 0.66$).} c) Evolution of the (normalized) mean of $\mS$ ($M=5000$) as a function of time for different values of the turnover rate $\tau$. The dashed line represents the threshold value divided by the smallest value of $\mS_{t=0}$ ($M=5000$) in the data. The data points were obtained from the datasets \cite{Chu2019} (yellow), \cite{Pogorelyy2018} (green) and \cite{Britanova2016} (orange). Different markers indicate different individuals.}}
  \label{fig:fig2_but_with_S}
\end{figure*}

\begin{figure*}[!htbp]
\begin{center}
\includegraphics[width=.8\linewidth]{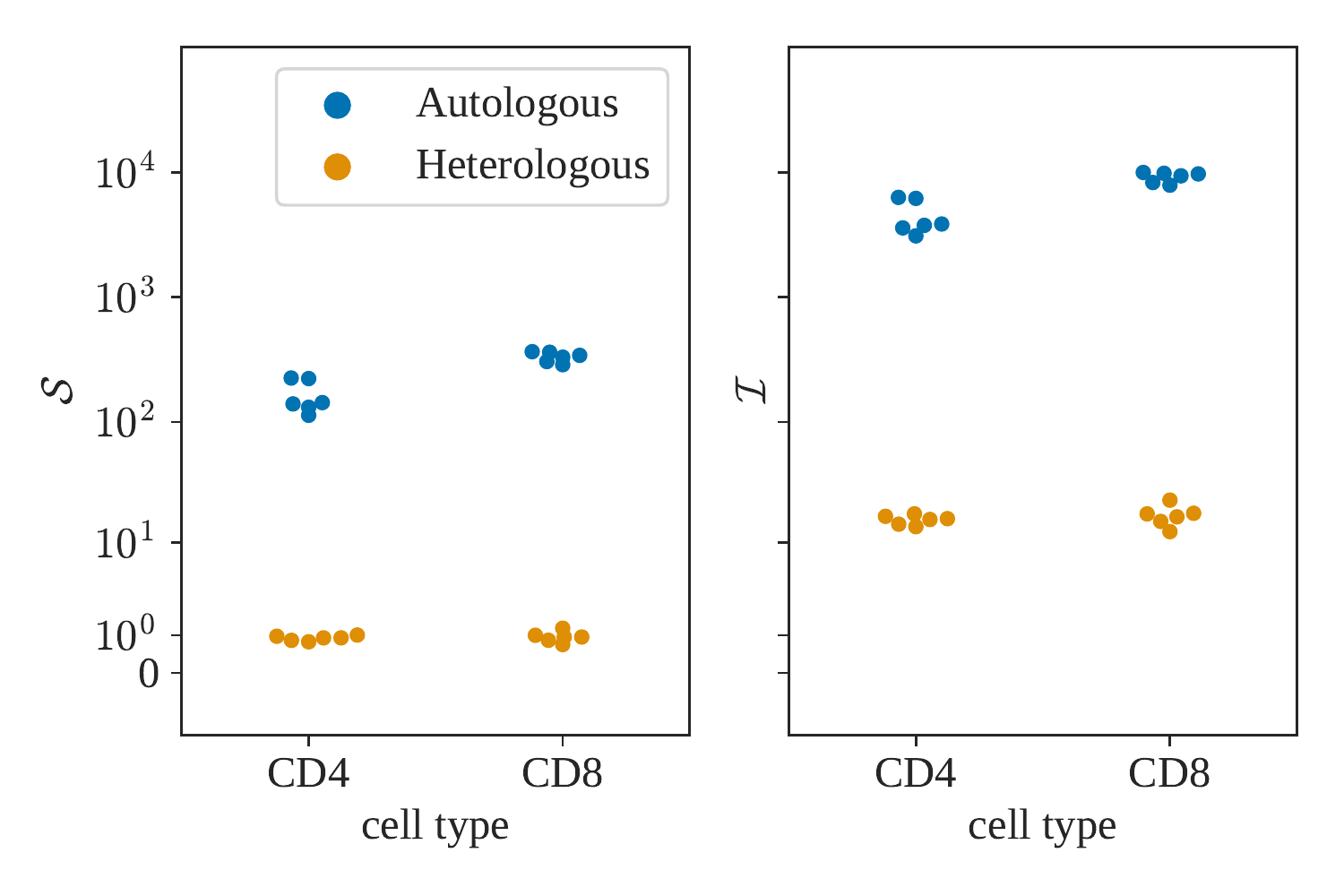}
\caption{\thomas{Distribution of $\mS$ (left) and $\mI$ (right) for CD4+ and CD8+ T-cells in the dataset from \cite{Pogorelyy2018} (6 individuals). In the autologous case (blue), there is a significant difference between the two cell types, caused by the differences between their clonal distributions.}
  \label{fig:CD4_CD8}
}
\end{center}
\end{figure*}

\begin{figure*}[!htbp]
\begin{center}
\includegraphics[width=1.\linewidth]{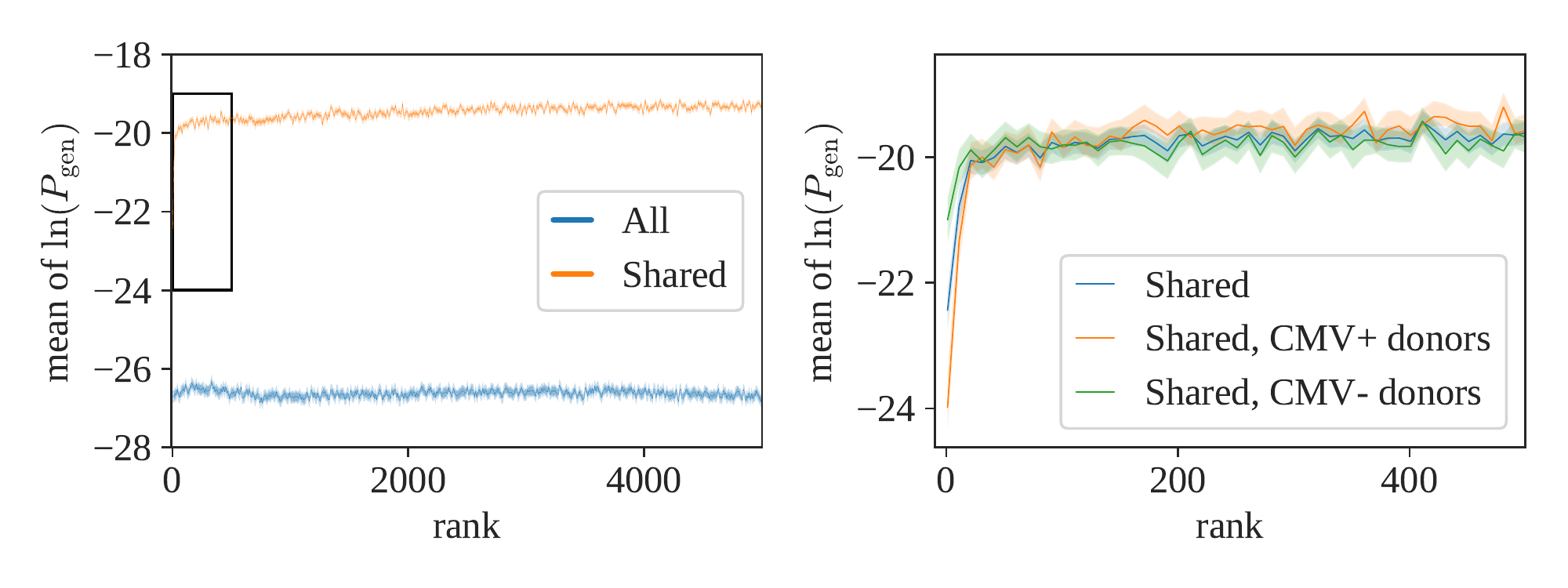}
\caption{Left: Mean value of $P_\mathrm{gen}$ as a function of the
  rank of the clonotype, for generic sequences (blue) and sequences
  shared between more than two donors (orange). The mean stays flat
  indicating that the probability of being generated does not
  generally depend on the clonotype size. There is an exception (black
  rectangle), shown
  as a close-up on the right panel. The top twenty clones, when shared between
  donors, have a smaller probability of being generated than expected
  by chance. This difference is likely
  to be driven by convergent selection against common pathogens, since CMV
  positive donors show a more prononced effect than CMV
  negative ones.
  \label{fig:rank_vs_pgen_plusplus}
}
\end{center}
\end{figure*}

\begin{figure*}[!htbp]
\begin{center}
\includegraphics[width=1.\linewidth]{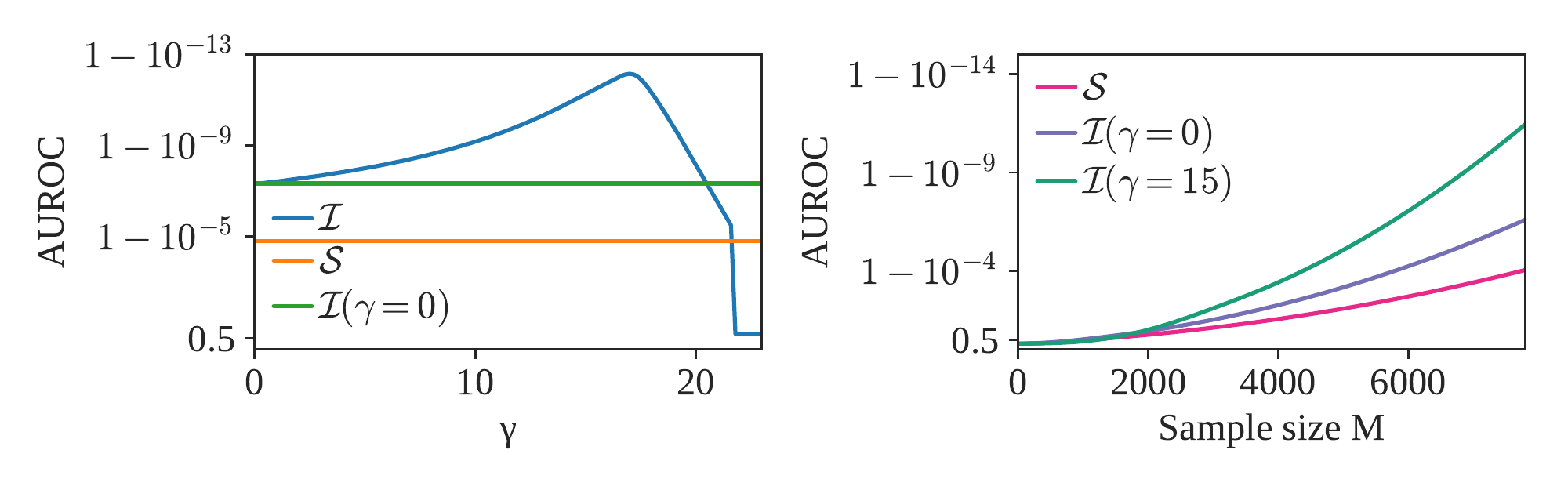}
\caption{Left panel: AUROC (Area Under Receiver Operating
  Characteristic) of $\mI$, as a function of $\gamma$
  ($M=M_1=M_2=5000$). We observe an optimum near $\gamma = 15$. Right panel:
  AUROC as a function of $M$, for $\mS$, $\mI(\gamma=0)$, and $\mI(\gamma=15)$.
\label{fig:Igamma}
}
\end{center}
\end{figure*}

\begin{figure*}[!htbp]
\begin{center}
\includegraphics[width=.8\linewidth]{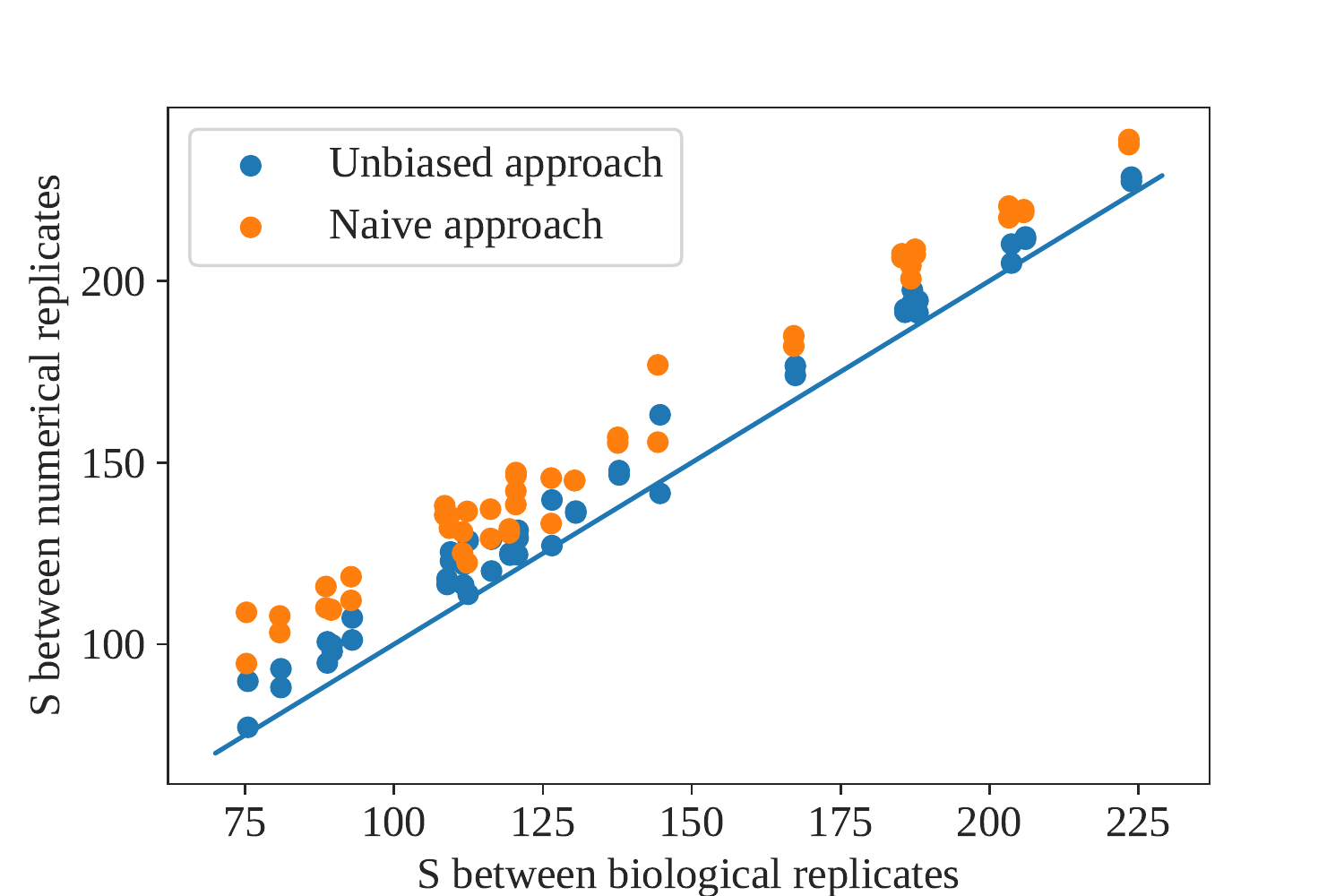}
\caption{Naive and
  corrected estimates of the autologous $\mS$ from single datasets,
  versus its values computed using true biological
  replicates from Ref.~\cite{Pogorelyy2018}.
\label{fig:S_unbiased_naive}
}
\end{center}
\end{figure*}

\begin{figure*}[!htbp]
\begin{center}
\includegraphics[width=.8\linewidth]{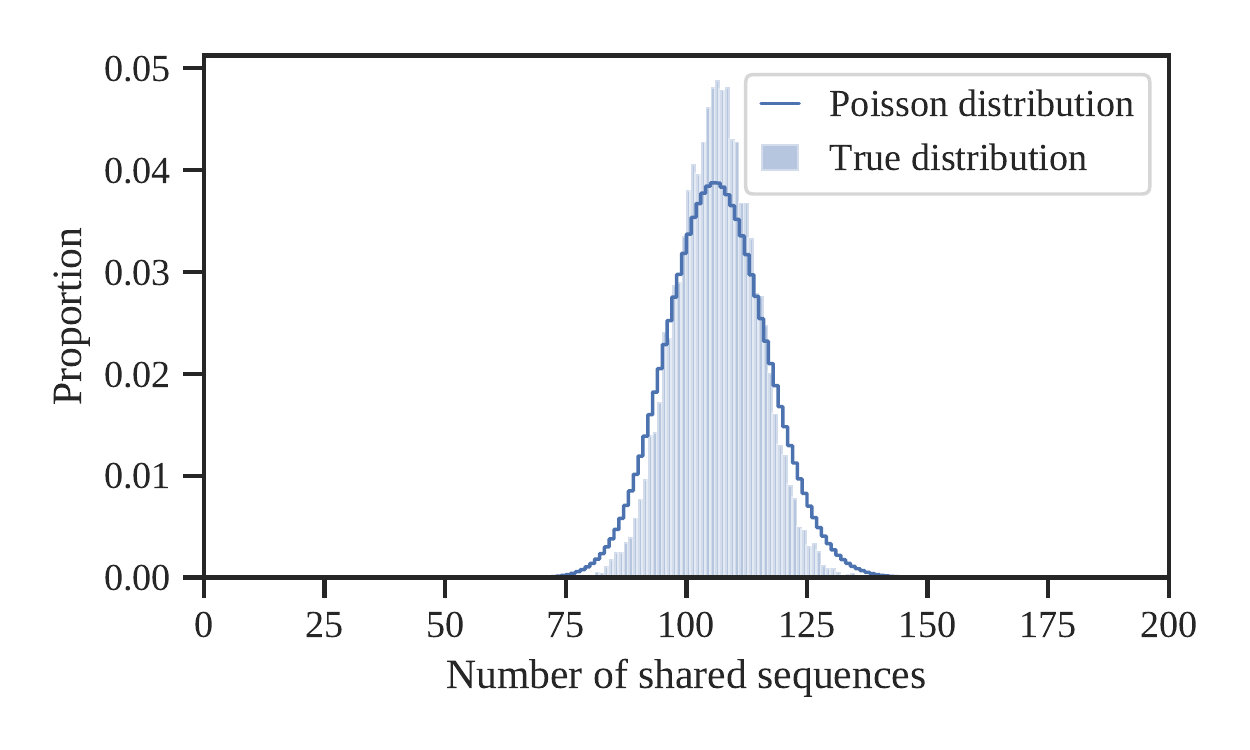}
\caption{Comparison between the distribution of $\mS$ obtained by
  computationally and repeatedly downsampling a single repertoire from
  Ref.~\cite{Emerson2017} with $M=5,000$ (histogram), and a Poisson distribution of the same mean (full line).
\label{fig:poisson_approx}
}
\end{center}
\end{figure*}

\begin{figure*}[!htbp]
\begin{center}
\includegraphics[width=.8\linewidth]{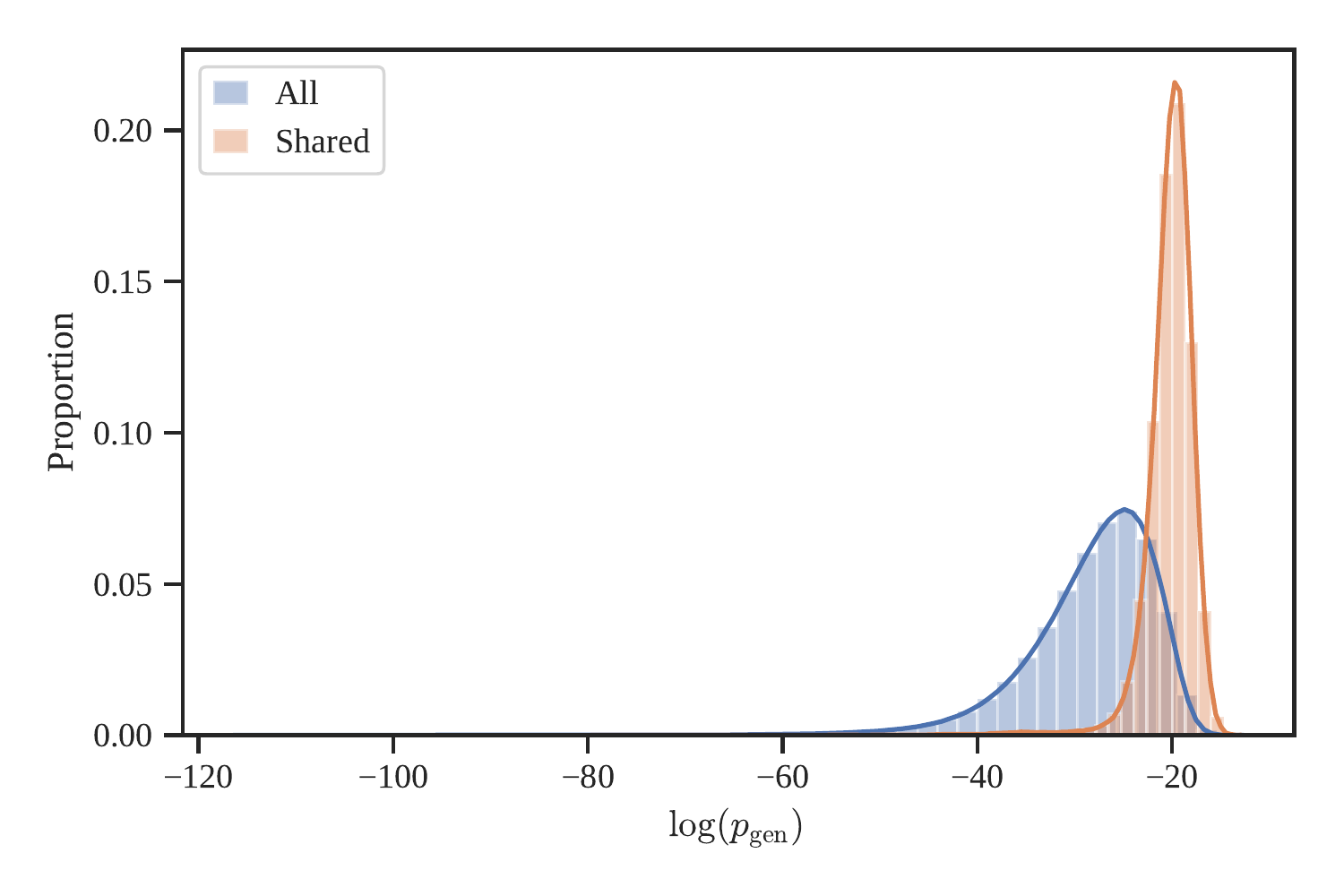}
\caption{Distribution of $\ln(P_{\mathrm{gen}})$ for generic
  sequences, and for sequences shared between heterologous samples.
\label{fig:pgen_distribution}
}
\end{center}
\end{figure*}

\end{document}